# Bayesian learning of adatom interactions from atomically-resolved imaging data

Mani Valleti[1], Qiang Zou[4], Rui Xue[5], Lukas Vlcek[2,3], Maxim Ziatdinov[4], Rama Vasudevan[4], Mingming Fu[4], Jiaqiang Yan[3], David Mandrus[5], Zheng Gai[4], and Sergei V. Kalinin[2,4]

[1] Bredesen Center for Interdisciplinary Research, University of Tennessee, Knoxville, TN 37996, USA

[2] Joint Institute for Computational Sciences, University of Tennessee, Knoxville, Oak Ridge, TN 37831, USA

[3] Materials Science and Technology Division, Oak Ridge National Laboratory, Oak Ridge TN 37831, USA

[4] The Center for Nanophase Materials Sciences, Oak Ridge National Laboratory, Oak Ridge, TN 37831

[5] Department of Materials Science and Engineering, University of Tennessee, Knoxville, TN 37996, USA

Atomic structures and adatom geometries of surfaces encode information about the thermodynamics and kinetics of the processes that lead to their formation, and which can be captured by a generative physical model. Here we develop a workflow based on a machine learning-based analysis of scanning tunneling microscopy images to reconstruct the atomic and adatom positions, and a Bayesian optimization procedure to minimize statistical distance between the chosen physical models and experimental observations. We optimize the parameters of a 2- and 3-parameter Ising model describing surface ordering and use the derived generative model to make predictions across the parameter space. For concentration dependence, we compare the predicted morphologies at different adatom concentrations with the dissimilar regions on the sample surfaces that serendipitously had different adatom concentrations. The proposed workflow is universal and can be used to reconstruct the thermodynamic models and associated uncertainties from the experimental observations of materials microstructures. The code used in the manuscript is available at https://github.com/saimani5/Adatom_interactions.

One of the key factors in understanding the physical functionalities and chemical reactivity of materials surfaces is the behavior of the adatom system, including the interactions between the adatoms and the substate, and interactions between adatoms, as well as associated local and global surface properties. Depending on the relative interaction energies, the adatoms can form surface gas and liquid phases, form multiple ordering types, and even give rise to the incommensurate surface phases.[1-9]

Traditionally, the properties of adatom systems were explored through the scattering methods such as low energy electron diffraction.[10-13] Here, the scattering pattern of reflected electrons yields insight into the surface and adatom configurations, much like conventional Laue scattering gives insight into crystal structure.[14] The quantitative analysis of the intensity-energy curves in LEED allows reconstruction of structural models, whereas high resolution LEED yields insight into long-range ordering mechanisms. However, the intrinsic difficulties in inversion of many-body scattering data and intrinsically low-k resolution of LEED has severely limited popularity of this approach, and in conjunction with advances in synchrotron light sources led to rapid proliferation of the surface-sensitive X-Ray scattering.[15-20] This average structural information has provided insight into the surface phases and their evolution that can be directly compared to the prediction of lattice or molecular dynamics models.

The harbinger of a new era in surface studies was the invention of Scanning Tunneling Microscopy (STM)[21, 22] and later non-contact Atomic Force Microscopy (nc-AFM).[23] In the decades since their invention, these techniques have evolved into highly robust methods for visualization of surface atomic structures, allowing localization of individual atoms and their groups.[24] This allowed scientists to gain insight into the structural and, via tunneling spectroscopy, electronic properties of the surfaces and adatom structures in unprecedented detail.[25-27]

However, the capability to visualize surface atomic structures have not yet fully been matched by development of analysis tools that can extract the physics of observed phenomena. Generally, such analysis necessitates several consecutive workflow tasks, including the transition from the images to materials specific descriptors, and subsequently recovering or building a correlative or generative model that can recreate the observed phenomena.

Here, we explore surface interactions in a system of Sulphur adatoms on CoSn terminated surface of a cleaved $Co_3Sn_2S_2$ crystal. We develop a machine learning workflow that allows seamless transformation of observed scanning tunneling microscopy images to atomic coordinates of surface and adatoms and referencing them to ideal lattice models. We further develop a Bayesian optimization-based approach that allows matching the experimental observables to a lattice Hamiltonian model, thus recovering a parsimonious generative physical model of this system.

As a model system, we explore S adatom features on top of CoSn subsurface of $Co_3Sn_2S_2$ single crystal, a newly discovered magnetic kagome-lattice Weyl semimetal from the Shandite family[28-33]. The Shandite family $A_3M_2X_2$ crystallizes in a rhombohedral structure, with a CoSn Kagome-lattice sandwiched by S and Sn layers [34]. Weyl semimetals, as a type of topological materials, possess a three-dimensional linear dispersion. They can be realized when time reversal and/or inversion symmetry are broken [28-33]. The ball and stick model of S adatoms on top of CoSn subsurface of $Co_3Sn_2S_2$ crystal is shown in Figure 1 (a). A large size STM image of one of the cleavage surfaces of $Co_3Sn_2S_2$ is shown in Fig. 1(b). Large amount of S adatoms scatter on top of flat hexagonal subsurface of CoSn, in many forms including monomers, dimers, long 1D adatom chains, or zig-zag chains. The density of the adatoms varies from area to area. Fig. 1(c) shows an atomic resolution zoomed-in STM image on one of the terraces of (b). Both the S adatoms (brighter atoms in chains and other irregular shapes) and hexagonal subsurface lattices are clearly visible. Some of the adatoms can be shifted or removed by scanning under harsher conditions. Fig. 1 (d) shows two images of the same areas before and after an imaging scan at 0.1A, -30mV. The red and blue dotted atoms in the blue and red circles are either moved to a nearby position or were removed from the surface.

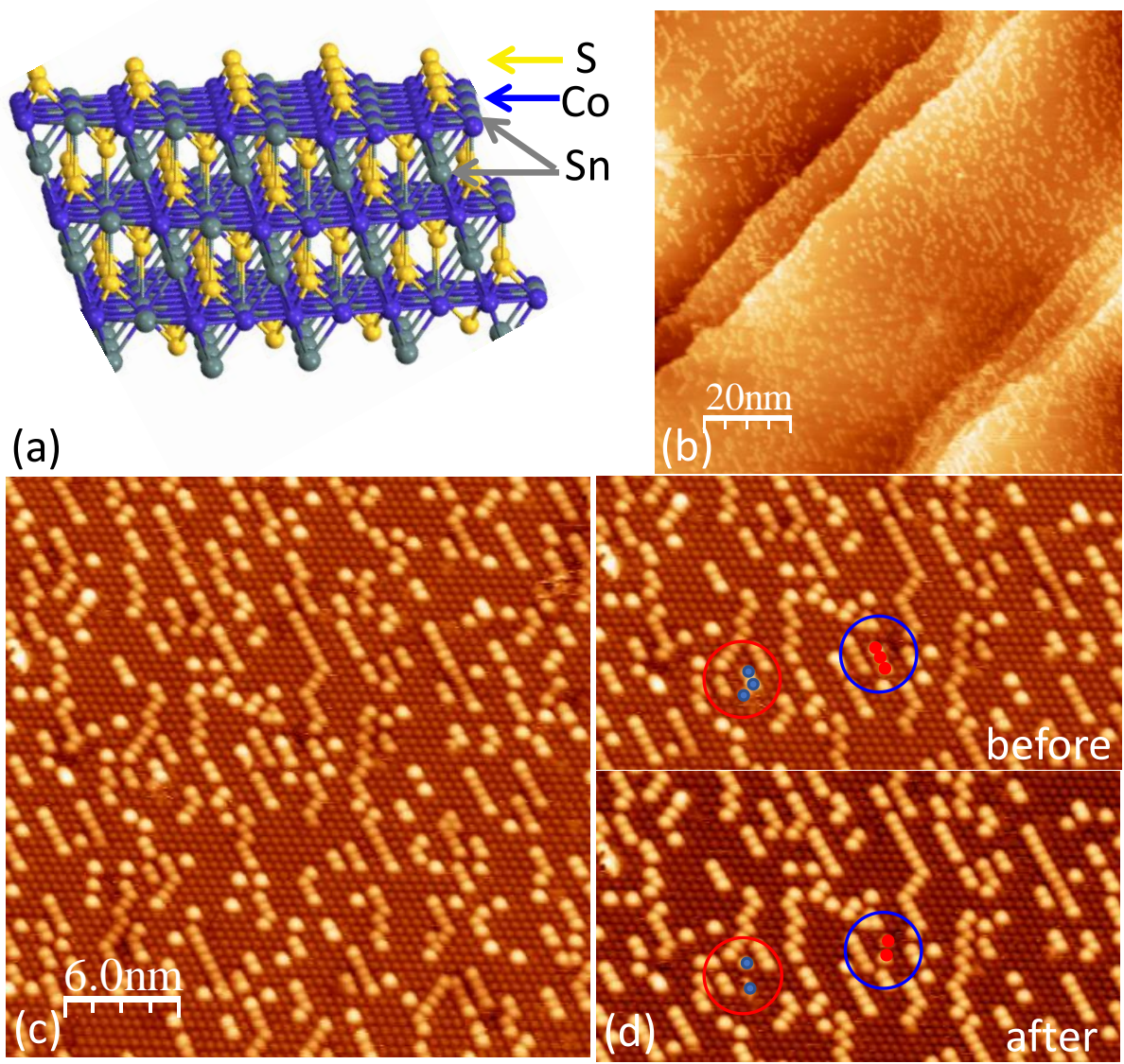

**Figure 1.** (a) Ball and stick model of $Co_3Sn_2S_2$ crystal with S adatoms on top of CoSn subsurface. (b) STM image of one of the cleavage surfaces of $Co_3Sn_2S_2$ (104 nm x 104 nm, 0.1 nA, -2000 mV)), shows large of S adatoms on top of flat subsurface of CoSn. (c) Zoomed in STM image on one of the terraces of (b) (30 nm x 30 nm, 0.1 nA, -100 mV)). (d) Some of the adatoms can be shifted or removed by scanning under harsher condition. Large red and blue circles outline two such areas before and after a 0.1A, -30mV scan.

The STM image in Fig. 1 (c) clearly illustrates the tendency of adatoms to form the elongated chains, breaking the $D_{3h}$ symmetry of the underlying surface. This symmetry breaking indicates strong anisotropic interactions between Sulphur atoms. Furthermore, the observed mobility of the atoms suggests that the adatoms are not trapped in a very deep potential wells, and hence the observed structures are close to being thermodynamically equilibrated locally. Here, we aim to construct models that can gain quantitative insight into this ordering behavior.

One such approach can be based on purely correlative models. In these models, the relative probabilities of first, second, and subsequent atomic neighborhoods are analyzed and can further be used to generate similar microstructures. This can be accomplished through dimensionality reduction methods such as principal component analysis, more complex strategies based on the standard or variational autoencoders, or development of a suitable low dimensional embedding such as *graph2vec* methods.[35, 36] However, these correlative models necessitate large volumes of data to train and do not offer direct physical insight into the observed behaviors. Furthermore, introducing progressively more complex descriptors will lead to severe data scarcity since the amount of available experimental data is highly limited.

An alternative approach is offered by the recovery of generative models. In this approach, it is assumed that the observed microstructures emerge as a result of time evolution of the model encoding physical interactions in the system (as opposed to all possible states), and the analysis seeks to recover the model parameters.[37-40] Note that in this context the model includes both the specific class and parameters within the class. Given the nature of the observed data that can be well-represented as partially occupied ideal adatom sublattice on the fully occupied surface atom lattice, here we use the triangular Ising model as a generative lattice model.

As the first step of the analysis, we seek to recover the coordinates of the surface adatoms in the lattice coordinates. To achieve this goal, we use blob detection class available in the scikit-image library.[41] The radii of the blobs detected is directly proportional to the standard deviation of the gaussian distribution used in finding the blobs. The surface atoms are then differentiated from the ad-atoms based on the standard deviation of the gaussian used to detect them. The coordinates of detected surface atoms and ad atoms are shown in Fig. 2 (b).

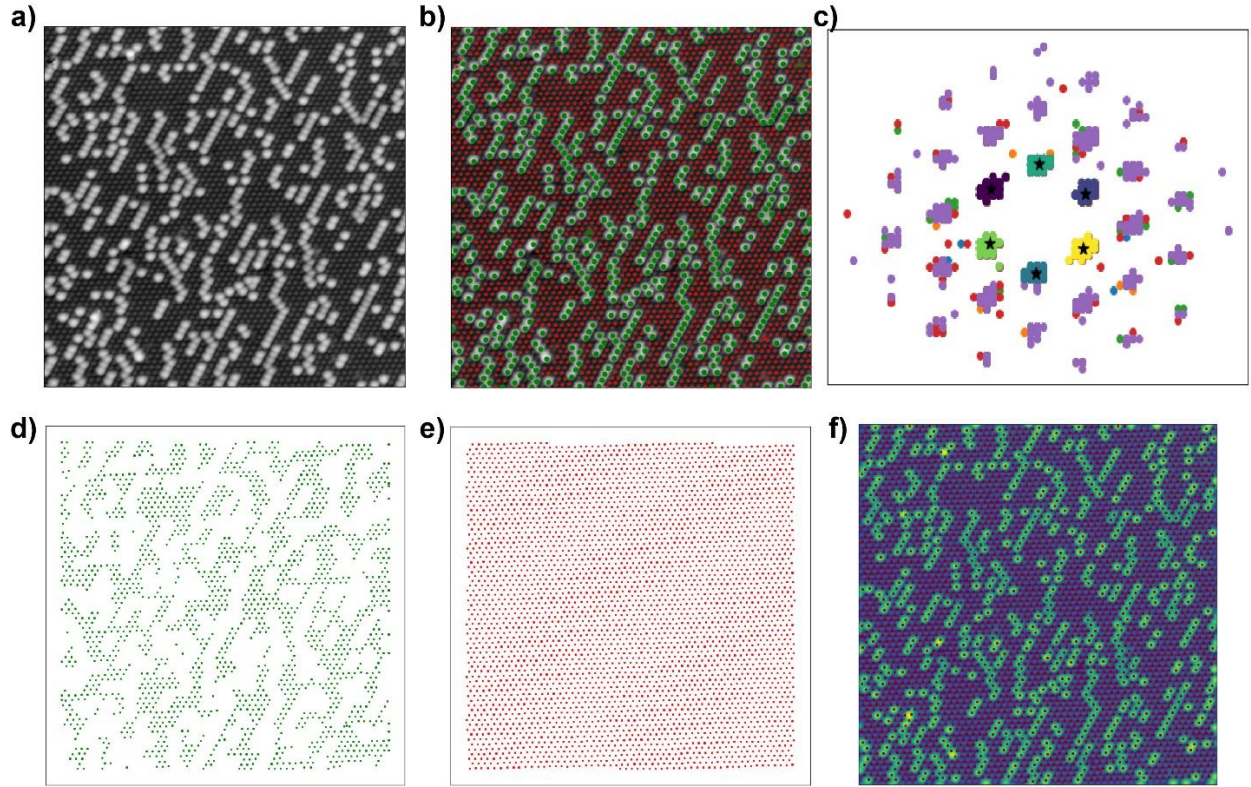

**Figure 2.** (a) STM image of the S adatoms on top CoSn surface and (b) identified surface atoms (red) and adatoms (green). (c) Nearest neighbor distribution of first six nearest neighbors corresponding to all the surface atoms detected (d) Observed surface atom lattice and (e) fully reconstructed surface lattice. (f) Overlay of the observed adatom lattice and reconstructed possible sites for ad atoms.

Some of the surface atoms are not visible in the image due to the overshadowing by the larger ad-atoms. To reconstruct the entire surface atom lattice, we first obtain the Bravais lattice corresponding to the surface atoms. This is done by plotting the positions of six nearest neighbors with respect to each surface atom. We then select the neighbors that are closer to the center than a certain threshold and divided them into six classes using k-means as shown in Fig 2 (c). The centers of these clusters are then the positions of surface atoms in the triangular Bravais lattice. The set of surface atoms detected using blob detection technique is shown in Fig. 2 (d) and the entire surface atom lattice is reconstructed using the obtained Bravais lattice and is shown in Fig. 2 (e). The possible sites for the adatoms are then derived from the fully reconstructed surface atom lattice and are shown along with the ad-atoms in Fig. 2 (f). This analysis allows us to determine both the lattice coordinates and real-space coordinates of the observed adatoms and observed surface atoms, as well as reconstruct likely positions of the unobserved surface atoms (shadowed under adatoms), providing thus a complete reconstruction of the surface crystal structure.

As a relevant descriptor that connects the observed surface structures and the modelling, we choose the relative frequencies of appearance of the atomic neighborhoods. This approach was derived by Vlcek[37, 42, 43] earlier and is dubbed statistical distance minimization. Here, we aim to minimize the statistical distance between the observations and the model, given as

$$s = \arccos\left(\sum_{i=1}^{k} \sqrt{p_i}\sqrt{q_i}\right) \tag{1}$$

Where $s$ is the statistical distance, a similarity measure between two distinct thermodynamic systems, $p_i$ and $q_i$ are the probabilities of configurations $i$ in the measurement of systems $P$ and $Q$, respectively, with the total of $k$ possible outcomes. This description was shown to be rigorous for the system in a state of thermodynamic equilibrium. For a triangular Bravais lattice, we collected the histograms corresponding to the six nearest neighbors.

As a generative model, we chose the Ising model on the triangular lattice, which reflects the observed structure of the adatom lattice. The Ising Hamiltonian is given by

$$H(\sigma) = -\sum_{<i,x>} J_{ix}\sigma_i\sigma_x - \sum_{<i,y>} J_{iy}\sigma_i\sigma_y - \sum_{<i,z>} J_{iz}\sigma_i\sigma_z \tag{2}$$

Where $H$ is the Hamiltonian of a given configuration $\sigma$, ($x$, $y$, $z$) corresponds to the sites in three 60° axes in $C_3$ symmetry, $i$ is the central atom, $J_{ij}$ is the interaction parameter corresponding to the sites $i$ and $j$ and the summation runs over all the nearest neighbor combinations. The vacant adatom sites are treated as -1 (downward) spin and the occupied ones are treated as +1 (upward) spin. Interactions between empty sites are ignored. Monte Carlo simulations are run on a triangular lattice model subjected to Kawasaki dynamics where the adatoms hop to a randomly selected empty site. A hop is accepted if the energy of the newly formed state is less than the original state else, the probability of hopping is then determined by the Boltzmann distribution and is given by equation 2.

$$P_\beta(\sigma_i) = \frac{e^{-\beta H(\sigma_i)}}{\sum_j e^{-\beta H(\sigma_j)}} \tag{2}$$

Where $\beta$ is the inverse temperature. Kawaski dynamics helps in conserving the number of ad-atoms in the system. The statistics of configurations generated by the Monte Carlo simulations are then compared with those extracted from the image by calculating their statistical distance. A total of 128 distinct combinations of spins corresponding to the six nearest neighbors in a triangular lattice are used as descriptors. The set of interaction parameters that yield the minimum statistical distance from the experimentally observed structures are the interaction parameters we seek.

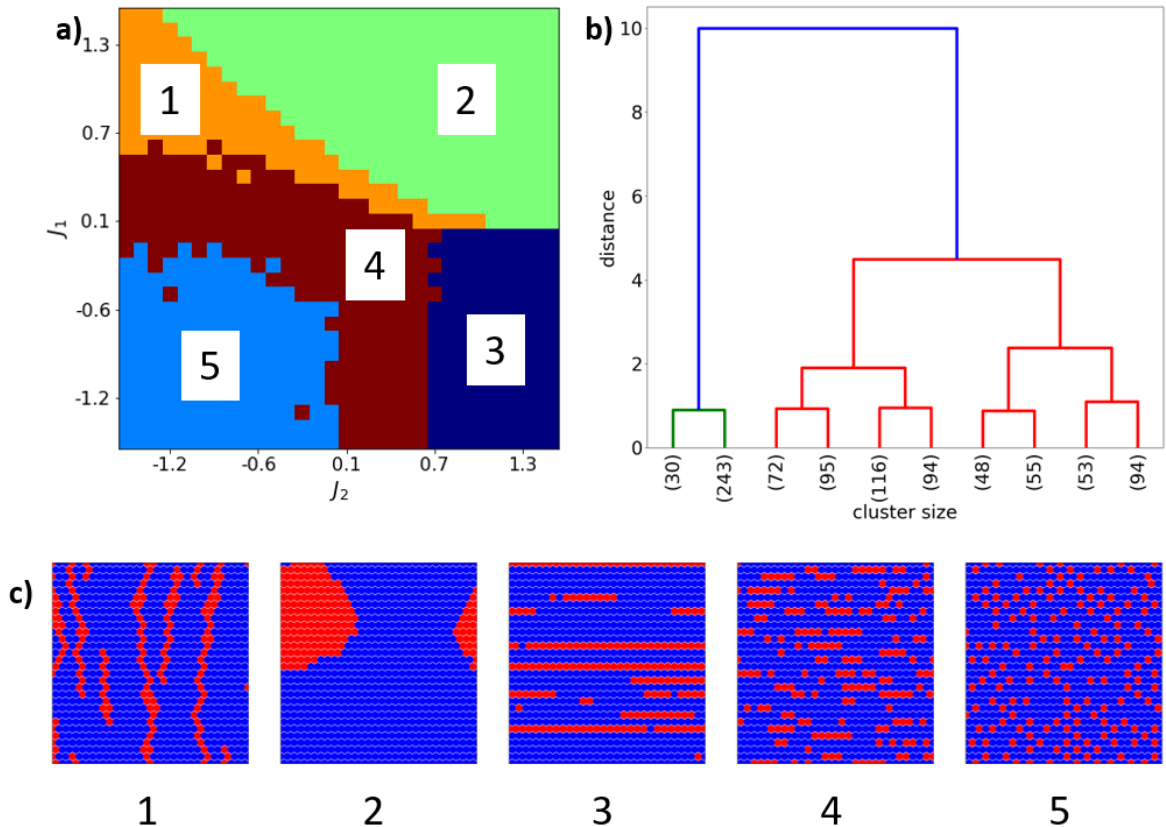

**Figure 3.** (a) Phase diagram of symmetric model ($J_x = J_y = J_1$ and $J_z = J_2$, $T_r = 0.8$) showing 5 distinct phases, (b) dendrogram showing the frequencies of classes, (c) Representative microstructures corresponding to the phases shown in (a).

Here, we consider both a symmetric model with two parameters $J_x = J_y = J_1$ and $J_z = J_2$, and the full model with all three integrals being non-equal. To obtain the phase diagram of the symmetric model, histograms are collected over the entire parameter space ($J_1 \in [-1.5, 1.5]$, $J_2 \in$ [1.5, 1.5]) and are then divided into five clusters using the k-means algorithm. The typical phase diagram for the symmetric model is shown in Fig. 3a and representative microstructures of each phase are shown in Fig. 3c. Relative class sizes and the distance between the classes are shown in the dendrogram in Fig. 3b. Depending on the relative signs of the interaction parameters, the phases corresponding to isolated adatoms (5), large-clusters (2), and ordered chains in one (3,4) and two possible directions (1) are realized. Note that these configurations can be expected from straightforward physical considerations, and here we did not aim to explore the phase diagram of the model Eq. (2) in detail. Rather we illustrate rapid mapping of configurations across a broad parameter space and associated qualitative morphological features. In fact, the proposed analysis pathway relies on the near-neighborhood configuration analysis, and hence is relatively insensitive to the long-range correlations, etc. that are traditionally explored in the context of statistical physics.

To extract the model parameters from the experimental data, here we develop an analysis based on the Bayesian optimization approach, an extension of the Gaussian Process (GP) regression towards guided search of parameter space.[44-46] Briefly, the GP regression refers to an approach towards interpolating, or learning, a function, $f$, given the set of observations $D = (x_1, y_1)$, ...$(x_N, y_N)$}. The arguments $x_i$ are assumed to be known exactly, whereas the observations are the sum of the function value and Gaussian noise with zero mean, $y_i = f(x_i) + \varepsilon$. The key assumption of the GP method is that the function $f$ has a prior distribution $f \sim \mathcal{GP}(0, K_f(x, x'))$, where $K_f$ is a covariance function (kernel).[47] The kernel function defines the relationship between the values of the function across the parameter space, and its functional form is postulated as a part of the fit. The learning is performed via Bayesian inference in a function space and the expected value of the function, corresponding uncertainties, and kernel hyperparameters are optimized simultaneously. The output of the GP process is then the predicted data set, uncertainty maps representing the quality of prediction, and kernel hyperparameters.

The aspect of the GP analysis that differentiates it from other interpolation methods is that not only the function value, but also associated uncertainty are determined over the parameter space. This can be used for effective exploration of the parameter space in an automated fashion. In this approach following a purely exploratory strategy, the subsequent measurement point is chosen as a region of maximal uncertainty of function $f$ after previous measurements. The GP method can be further extended towards Bayesian optimization, where the selection of the next measurement point in the parameter space is based both on the uncertainty and target value of the function.

Here we implement the Bayesian Optimization based on the GPyTorch library[48] and GPim package.[49] We used the RBF or Matern kernels, defined as

$$k_{RBF}(x_1, x_2) = \sigma^2 \exp\left(-0.5 \times \frac{|x_1-x_2|^2}{l^2}\right) \quad (2)$$

where $l$ and $\sigma^2$ are kernel length scale and variance, respectively, which are learned from the data by maximizing the log-marginal likelihood. The acquisition function, which determines the trajectory of subsequent measurements, was chosen to switch between pure exploration and exploitation with 60% probability towards exploitation. For exploitation, we aimed to minimize statistical distance between the experimental histogram and the histograms obtained from the Monte Carlo simulations.

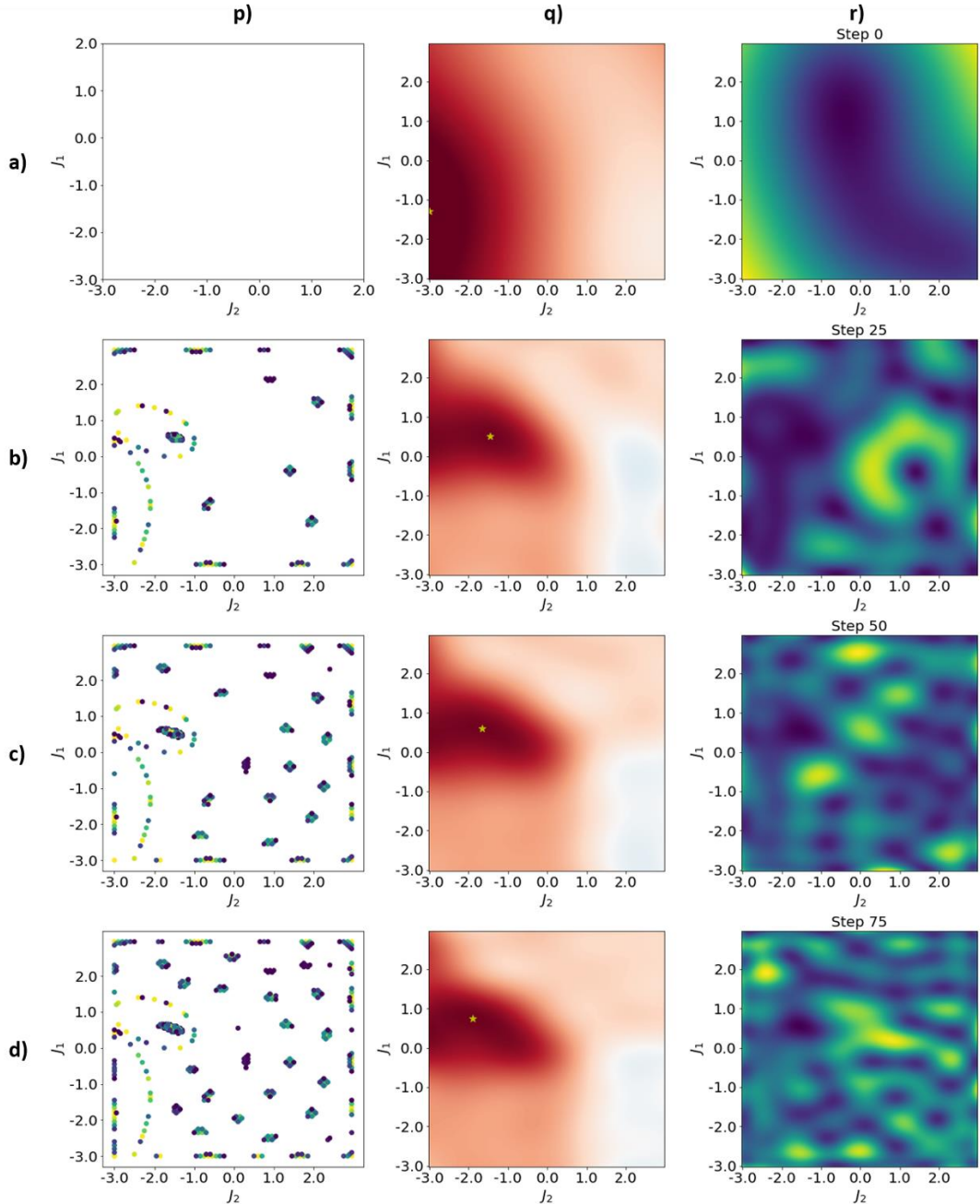

**Figure 4.** (p) Data points explored, (q) Statistical distance surface, (r) Uncertainty in statistical distance surface at the end of (a) Step-0, (b) Step-25, (c) Step-50, (d) Step-75. The yellow cross indicates the combination of parameters that yield the minimum statistical distance after the corresponding step.

The flow of the Bayesian optimization is delineated in Fig. 4. Here, shown is the explored points in the parameter space, the reconstructed statistical distance, and uncertainty. Outcomes at the end of Bayesian optimization step – 0, 25, 50, 75 are shown in Fig. 5a-d respectively. Data points explored, interpolated statistical distance surface and uncertainty in the interpolation for the different steps are shown in Fig. 5p-r. Since the shape of the statistical distance surface is simple with no multiple local minima, the region of global minimum is identified in the first few steps. 25 steps are sufficient to generally delineate the position of the minimal statistical distance, corresponding to the maximal (in a sense of thermodynamics of the generative model) similarity between the experimental data and realizations of the generative model. This saves a lot of computational power when compared to doing a grid search over the entire space. To determine the reproducibility, we run analysis 10 times and the determined interactions parameters are $J_1 = 0.59$, $J_2 = -1.79$ at $T_r = 4.0$.

To attest the veracity and systematic error of the proposed approach, we further explore to which extent the known interaction parameters of the symmetric model can be reconstructed from experimentally observed histograms. As an example, we collected the statistics of two distinct pseudo-experimental cases ($J_1 = 0.5$, $J_2 = -0.5$, $T_r = 4.0$ and $J_1 = 1.0$, $J_2 = -1.0$, $T_r = 0.8$). The purpose of the pseudo experimental simulations is that their statistics mimic the real-world data. The proportion of the adatoms in these simulations are consistent with the experimental data. The histograms of the pseudo experimental cases are then compared with the simulations over the entire parameter space at the same reduced temperature using statistical distance. At stronger interactions (lower temperature and/or higher values of interaction parameters), the statistical distance minimization technique identifies the phase of the experimental data.

These reconstructions are illustrated in Fig. 5, where we try to reconstruct the interaction parameters of case-2 ($J_1 = 1.0$, $J_2 = -1.0$, $T_r = 0.8$). The microstructures corresponding to this case are shown in Fig. 5 (d), where clearly the clusters of adatoms are visible, suggesting a 'pre-transition' region. Around the transition region, reconstruction would be ideal with minimum uncertainty and the uncertainty/shallowness of the statistical distance surface increases with increase in temperature. This effect can be observed in the reconstruction of case – 1 in Fig. 5 (a) and the corresponding microstructures are shown in Fig. 5 (c). Shallowness of the statistical distance surface can also be attributed to the low concentration of adatoms present.

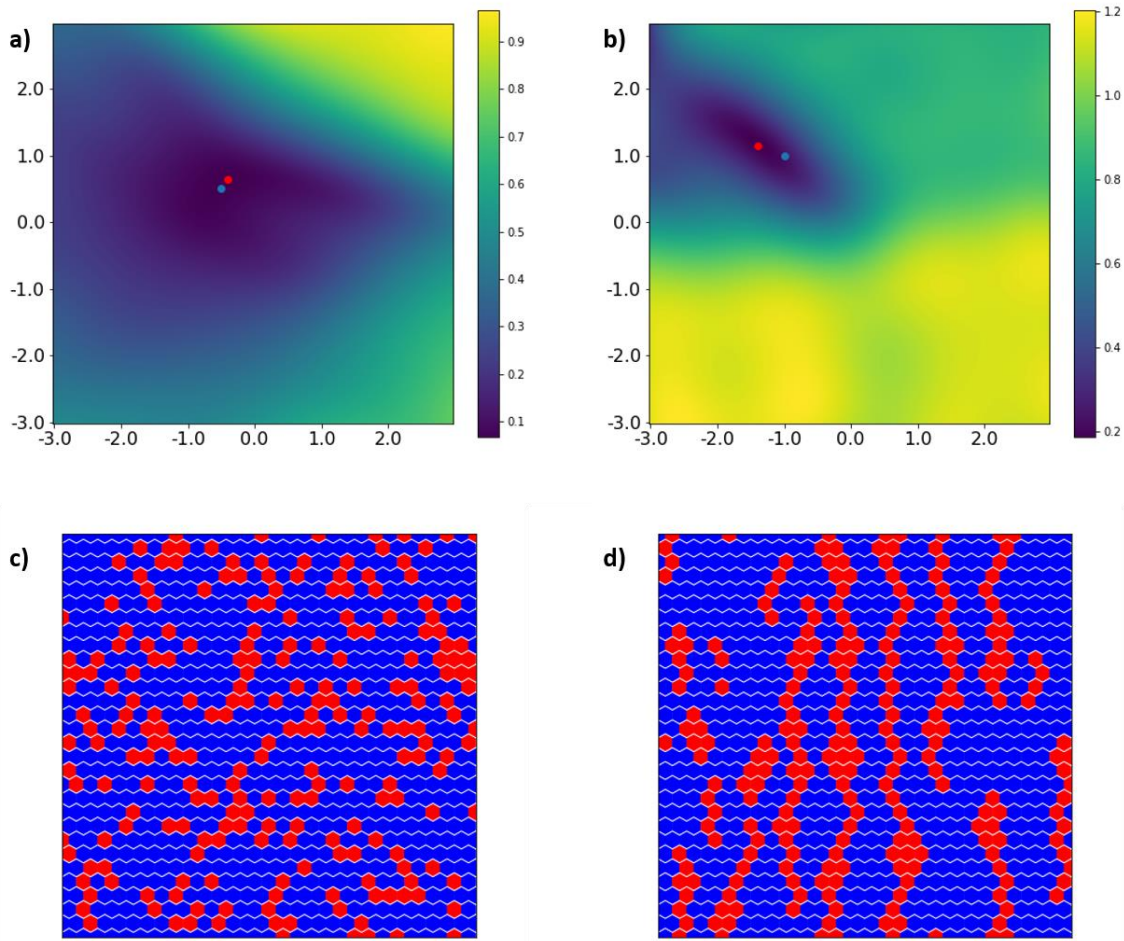

**Figure 5.** Statistical distance as a function of interaction parameters for (a) case - 1 ($J_1 = 0.5$, $J_2 = -0.5$, $T_r = 4.0$) and (b) case - 2 $J_1 = 1.0$, $J_2 = -1.0$, $T_r = 0.8$), Representative microstructures for (c) case - 1 and (d) case - 2. Red and blue sites correspond to occupied and empty sites respectively.

However, the analysis clearly suggests that the observed morphologies of the adatom structures allows to establish the nature of the adatom interactions (attractive in 2 directions, repulsive in one direction), and establish the approximate values of interaction parameters. The characteristic morphology of the error surface suggests that the ratio of the parameters can be determined reliably, whereas absolute value is associated with larger errors.

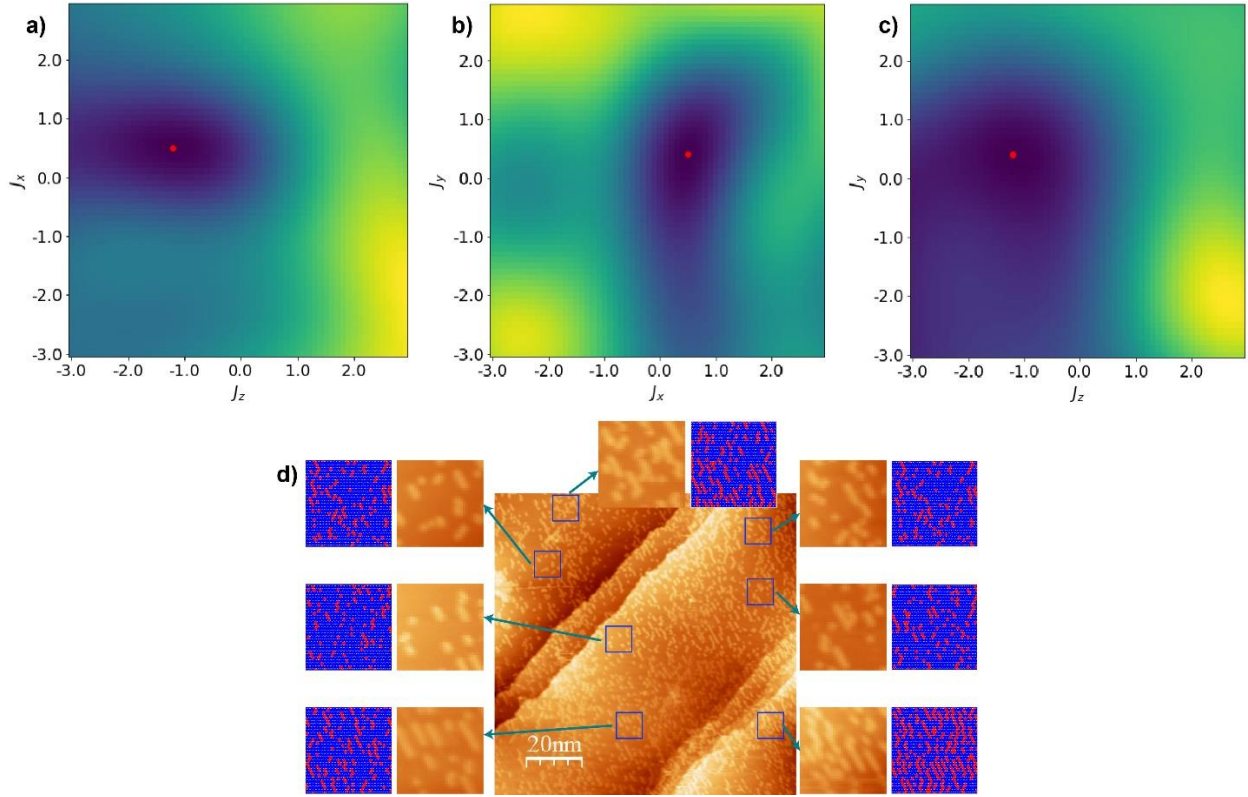

**Figure 6.** Statistical distance landscapes of the 3D model for optimal values of (a) $J_y = 0.4$, (b) $J_z = -1.2$ and (c) $J_x = 0.5$. The red dots indicate the combination of optimal parameters after projecting the statistical distance curve onto a 2D surface. (d) Randomly selected sites with different concentrations of adatoms and the predicted configurations using the optimal values for interaction parameters.

We further apply the technique discussed to a 3D parameter space where $J_x$ and $J_y$ are treated differently. The set of optimal values provided by the analysis are $J_x = 0.5$, $J_y = 0.4$, $J_z = -1.2$ in the units of $k_BT$. The statistical distance surface formed is then projected onto 2D surfaces for visualization. The 2D surfaces selected are the optimal values of the each of the interaction parameters. Statistical distance landscapes after projection onto the optimal values of the interaction parameters are shown in Fig. 6 (a-c). Statistical distance surface in 3D is projected onto $J_x$-$J_z$ (Fig. 6a), $J_y$-$J_x$ (Fig.6b), and $J_y$-$J_z$ and (Fig. 6c) at the optimal values of $J_y = 0.4$, $J_z = -1.2$ and $J_x = 0.5$ respectively. Several random areas from the image are then selected with varying concentrations of adatoms and the reconstructed configurations obtained with the optimal values of interaction parameters are shown for comparison (Fig. 6d).

To summarize, here we explored the surface structures of adatoms on the shandite surface. The adatoms tend to form 1D chains breaking the hexagonal symmetry of the surface and are weakly mobile under observation conditions. This suggests the attractive interactions between the atoms, in agreement with the known chemistry of sulphur favoring chain-like molecule formation.

To extract the generative physical model, we use the Bayesian optimization to minimize statistical distance between the chosen physical model and experimental observations. As a physical model, here we explore the 2- and 3-parameter Ising models on triangular lattice with the particle density corresponding to experimental adatom density. It was shown that the cost surface corresponding to the experimentally relevant parameter space region has an elongated minimum, resulting in the relatively large errors of reconstruction both for synthetic (known ground truth) and experimental data. With this limitation, the interaction parameters of $J_1 = 0.59$, $J_2 = -1.79$ for 2D and $J_x = 0.5$, $J_y = 0.4$, $J_z = -1.2$ for 3D are obtained.

Thus, the derived generative model in turn allows generalization across the parameter space. For concentration dependence, we compare the predicted morphologies for different adatom concentrations with the dissimilar regions on the sample surfaces that serendipitously had different adatom concentrations.

We believe that the proposed workflow is universal and can be used to reconstruct the thermodynamic models and associated uncertainties from scarce experimental observations of materials microstructures at a limited computational cost. We note that critical consideration going further will be the analysis of the non-equilibrium effects, i.e. presence of frozen interactions. Here, statistical distance minimization requires proximity to (local) thermodynamic equilibrium, whereas observability via imaging requires the frozen atomic configurations. However, overall, the proposed approach opens the pathway for extracting generative models from observations.

**Acknowledgements:** This effort (, feature extraction) is based upon work supported by the U.S. Department of Energy (DOE), Office of Science, Basic Energy Sciences (BES), Materials Sciences and Engineering Division (S.M.V., S.V.K., R.K.V.), and scanning tunneling microscopy was conducted at the Center for Nanophase Materials Sciences (CNMS), a U.S. Department of Energy, Office of Science User Facility.

**Materials and methods**

**Synthesis:** $Co_3Sn_2S_2$ crystals were synthesized using the self-flux method using a procedure similar to that described in Ref[50]. Co slugs(Alfa Aesar, 99.995%), Sulfur pieces(Alfa Aesar, 99.9995%) and Sn shots(Alfa Aesar, 99.99+%) with an atomic ratio of Co:S:Sn=9:8:83 were placed in a 2 ml $Al_2O_3$ Canfield crucible set[51] and sealed into a silica tube under vacuum. The tube was heated to 400°C at 100°C/h. After dwelling for 4 hours, the tube was heated with the same rate to 1100°C and kept at this temperature for 24 hours. The tube was then cooled to 700°C at 3°C/h prior to separating the flux from the crystals in a centrifuge. After centrifuging at 973K to separate the crystals from the flux, single crystals were obtained from the crucibles.

**Characterization:** Magnetic measurements were performed using a Quantum Design Magnetic Property Measurement System (MPMS) which has Reciprocating Sample Option (RSO) and ac susceptibility options. Phase purity, crystallinity, and the atomic occupancy of all crystals were checked by collecting Powder X-ray diffraction (XRD) data).

**STM and STS:** Crystals were cleaved in ultra-high vacuum (UHV) at ~ 78 K and then immediately transferred to the Scanning Tunneling Microscopy/Spectroscopy (STM/S) head which was precooled to 4.2 K or 78 K without breaking the vacuum. The STM/S experiments were carried out at 4.2 K or 78 K using a UHV low-temperature and high field scanning tunneling microscope with base pressure better than $2\times10^{-10}$ Torr. Pt-Ir tips were mechanically cut then conditioned on clean Au (111) and checked using the topography, surface state and work function of Au (111) before each measurement. The STM/S were controlled by a SPECS Nanonis control system. Topographic images were acquired in constant current mode with bias voltage applied to sample, and tip grounded. All the spectroscopies were obtained using the lock-in technique with a modulation of 0.1 to 1 mV at 973 Hz on bias voltage, dI/dV.